\documentclass[aps,preprintnumbers,nofootinbib,superscriptaddress,twocolumn]{revtex4}

\usepackage{graphicx}
\usepackage{epsfig,graphics}
\usepackage{amsfonts,amssymb,amsmath}
\usepackage{amsthm}

\newcommand{\comment}[1]{}

\theoremstyle{plain}

\theoremstyle{definition}

\begin{document}

\title{Location-Oblivious Data Transfer with Flying Entangled Qudits}

\author{Adrian \surname{Kent}}
\affiliation{Centre for Quantum Information and Foundations, DAMTP, Centre for
  Mathematical Sciences, University of Cambridge, Wilberforce Road,
  Cambridge, CB3 0WA, U.K.}
\affiliation{Perimeter Institute for Theoretical Physics, 31 Caroline Street North, Waterloo, ON N2L 2Y5, Canada.}

\date{\today}

\begin{abstract}
We present a simple and practical quantum protocol involving two
mistrustful agencies in Minkowski space, which allows Alice to
transfer data to Bob at a spacetime location that neither can 
predict in advance.  
The location depends on both Alice's and Bob's actions.  The protocol
guarantees unconditionally to Alice that Bob learns
the data at a randomly determined location; it guarantees to Bob that Alice will not learn
the transfer location even after the protocol is complete. 

The task implemented, transferring data at a space-time location that remains
hidden from the transferrer, has no precise analogue in non-relativistic quantum cryptography.   
It illustrates further the scope for novel cryptographic applications
of relativistic quantum theory.   
\end{abstract}

\maketitle

\section{Introduction}
\label{sec:intro}

Because quantum states cannot be cloned, and can be entangled, 
someone who generates quantum states can retain more information
about them than another party who receives them.   Quantum
cryptography exploits this information control (and other
features of quantum theory) to implement interesting cryptographic
tasks with security guaranteed so long as quantum theory is correct.

The no-signalling principle of special relativity also allows
the controlled distribution of information, in the sense that 
a party can transfer classical or quantum information secure
in the knowledge that it can only be communicated to the future
light cone of the transfer point.   The important cryptographic
task of mistrustful coin tossing \cite{cointoss} can be implemented
using this fact alone.    More surprisingly, it can also be used
to implement the (strictly more powerful \cite{power} )
task of bit commitment, via protocols that require only classical information
exchanges but which guarantee security against Mayers-Lo-Chau quantum
attacks \cite{mayersvthree,mayersprl,mayersone, lochauprl,lochau} 
as well as all classical attacks \cite{kentrel,kentrelfinite}. 

We would like, for both theoretical and practical reasons, 
to understand precisely which cryptographic tasks 
can be implemented when we take relativistic quantum theory to be 
the underlying theory, and thus combine the information control techniques allowed by
both theories.   There have been several interesting recent
illustrations \cite{colbeckkent,bcsummoning} of the cryptographic 
power of relativistic quantum theory.   Work on quantum tagging and position-based
quantum cryptography has also produced new cryptographic schemes for quantum
tagging secure against technologically restricted adversaries \cite{taggingpatent,malaney,chandranetal,kms}, 
some possible limits to the
level of security attainable for some versions of these tasks \cite{kms,chandranetal}
and an unconditionally secure protocol \cite{kenttaggingcrypto} for quantum tagging 
given a tagging device that is cryptographically secure -- i.e. that can keep internal data secret from
an adversary.   Intriguing extensions of quantum tagging to more general position-based cryptographic
tasks have also been developed \cite{chandranetal}.  

Light speed constraints are already practically significant 
in terrestrial technology, even on small scales, and will become far more so as we
and our technology migrate further beyond the Earth.
So it seems natural not only to try to apply these constraints to implement classical (non-quantum
non-relativistic) tasks, but also to rethink what cryptography actually means -- what the most 
general interesting and potentially useful cryptographic tasks are -- in a world governed by 
relativity and quantum theory.   
Although Shor's and Grover's algorithms, and other efficient quantum algorithms for intrinsically
classical computational tasks, were seminal discoveries which added great impetus to the development of
quantum computational theory, many believe 
that intrinsically quantum computational tasks may well 
ultimately be the most significant applications of quantum computing.
It seems by now quite plausible too that {\it intrinsically} relativistic and quantum cryptographic 
tasks will also ultimately be among the most significant applications of physics-based 
cryptography and computing.   At the very least, it seems worth exploring what might be possible
with this vision in mind.   

In this paper we define a new and intrinsically relativistic cryptographic
task, data transfer at a spacetime location that remains hidden from the
transferring party.   We give a simple and practical protocol for
implementing this task, and show the protocol is unconditionally
secure, in an appropriately defined sense.
The protocol is inspired by the 
no-summoning theorem \cite{nosummoning} and its application 
to bit commitment \cite{bcsummoning}. 
Roughly speaking, the underlying intuition is that, by producing one qudit of an entangled pair
at an unpredictable point, Alice effectively constrains Bob, who has no way of ensuring
that he can produce the other qudit at or near the same point.   
However, the security proof goes beyond the no-summoning theorem,
using other features of relativistic quantum theory.  In particular, it 
implicitly assumes that the unitary evolution of quantum states between any
two spacelike hypersurfaces can be covariantly defined, as in
the Tomonaga-Schwinger formalism \cite{tomonaga, schwinger}.  

\section{Transferring Data with Entangled Flying Qudits}

We give an idealized version of the protocol here to simplify the 
presentation.\footnote{As in other applications \cite{bcsummoning}
inspired by the no-summoning theorem \cite{nosummoning}, these unphysical idealizations can be 
relaxed without affecting the
essential idea, and it is easy to adjust the protocol so that it can be securely implemented under realistic physical assumptions, allowing for noise and time lags.  See Refs. \cite{nosummoning, bcsummoning} for illustrations.}
We suppose that space-time is 
Minkowski and that nature is described by some appropriate relativistic version 
of quantum theory.   We consider two agencies, Alice and Bob, with representatives
at all the relevant points in space-time, and suppose that both 
have arbitrarily efficient technology, limited
only by physical principles, and that all their operations and communications
 are error-free.    
We also suppose they agree in advance on some
space-time point $P$, where the protocol commences, and on a fixed inertial reference 
frame. 

We suppose they are also independently able to access 
suitable points $Q_i$ (where $i$ runs from $0$ to $m-1$) in the causal future of $P$, and that each is able to keep
information everywhere secure from the other unless and until they
choose to disclose it.
We also suppose that Alice and Bob can independently and securely access 
all the points $P$ and $Q_i$ and instantaneously exchange information there.
In particular, Alice can keep a bipartite entangled state private
somewhere in the past of $P$ and arrange to transfer one 
subsystem to Bob at $P$, and Alice's and Bob's subsystems can then 
be kept private unless and until they choose to transfer them
at some point(s) in the future of $P$.        
We also suppose that both parties can send any relevant states
at light speed in prescribed directions along secure quantum
channels.   

Apart from our idealized assumptions about independent access to
suitable space-time points, we are following here the standard
conventions
of mistrustful cryptography.   Alice and Bob mistrust one another, and 
each trusts only operations they carry out themselves in their own secure
``laboratories''.  As is standard in mistrustful cryptography, the aim
of our protocol is to offer each some guaranteed control over the
information obtained by the other, despite their mistrust.   
We need not consider the actions of any third parties (for instance
an interfering eavesdropper Eve), since as far as Alice is concerned 
everything outside her laboratory is potentially under Bob's control,
and vice versa.  

We focus here on a specific version of the protocol, with $m=2$: the protocol can obviously be extended
to larger $m$, to various geometries, and to variants involving multipartite entanglement.   
For $m=2$, Alice and Bob agree two distinct spatial directions $v_0$ and $v_1$ in the agreed frame.
For simplicity, we focus here on one natural choice, in which the $v_i$ are opposite; we set $c=1$,
take $P$ to be the origin and the two spatial directions to be defined
by the vectors $v_0 = (-1 , 0 , 0, 0)$ and $v_1 = (1,0,0,0 )$  in the fixed frame
coordinates $(x,y,z,t)$.   

Before the protocol, Alice generates a two qudit state; the dimension
$d$ is a pre-agreed parameter.   This state determines the datum
that will be transferred in the protocol.   If the protocol is
followed, Bob will receive an integer in the range $1 , \ldots , d^2$.
Suppose for the moment that Alice wishes this integer to be
classically determined from the start of the protocol. 
She then prepares a maximally entangled state $\psi_i \in {\cal C}^d_A \otimes {\cal C}^d_B$, 
chosen from a pre-agreed orthonormal basis labelled by $i = 1 , \ldots , d^2$. 
This is encoded in two physical subsystems which (idealizing again) we take to be pointlike.  
She keeps the state private until $P$, where she gives the second subsystem to 
Bob.    Alice now uses a private equiprobable random bit $ j \in \{ 0 , 1 \}$ 
and sends the first subsystem along a 
secure channel that she controls at light speed in the direction $v_j$, i.e. along the line
$L_0 = \{ (-t, 0, 0, t) , t > 0 \}$ (for $j=0$) or the
line $L_1 = \{ (t, 0, 0, t) , t > 0 \}$  (for $j=1$).\footnote{By requiring light speed transmission in opposite
directions we ensure the possible data transfer points separate as fast as possible in the given frame.
Neither constraint is necessary, however: the protocol achieves location-oblivious data transfer so long as 
$Q_0$ and $Q_1$ are spacelike separated.}

Bob is free to act as he wishes, depending on his preferred distribution for the point of data
transfer.   If he wishes to maximize his chance of obtaining the data at the
earliest\footnote{With respect to the Minkowski causal partial ordering.} possible
point in space-time, then he also randomly chooses a bit $ k \in \{ 0 , 1 \}$ by flipping
a fair coin and sends the second subsystem along a secure channel that he controls at light speed in the direction $v_k$.  

To transfer the datum to Bob, Alice gives him the second subsystem at some point $Q_j$ along
the line $L_j$.   The points $Q_0$ and $Q_1$ (but not Alice's choice $j$) could be agreed
in advance, in which case Bob knows to expect the data at one of two possible points.
However, if Alice wishes to retain freedom in choosing the transfer point, she can
choose any point $Q_j$ along $L_j$, provided that Bob is prepared to receive data
there.    To simplify here, we assume $Q_0 = \{ -T, 0 , 0 , T \}$ and
$Q_1 = \{ T , 0 , 0, T \}$ are fixed, for some agreed choice of $T$.  

The transferred datum is the value of $i$.    To obtain this information, Bob needs to 
carry out a measurement of the two subsystems in the agreed orthonormal basis.
His  measurement need not necessarily be local; however, if not, then Bob will possess the datum $i$
only at points where he is able to combine all the data obtained from non-local 
measurement operations.   

\subsection{Security against Alice}

Alice has little scope for cheating.   She can choose the joint state of the two subsystems
to be a state other than one of the basis states $\psi_i$.   If she does, Bob will ultimately obtain
an $i$ corresponding to the random outcome of his measurement.   Nonetheless, Bob obtains
a value of $i$.   The aim of this protocol is simply to ensure that Alice transfers some datum to Bob: as with
classical oblivious transfer \cite{rabinot}, there is no requirement that the datum transferred should be 
``correct'' by any external standard.   Nor do we require that the datum be classically determined
before the protocol is complete -- something which is not possible with any intrinsically
quantum protocol (see e.g. \cite{kentshort}).   Alice's freedom in choosing the joint state thus does
not constitute cheating.   

Alice's only other freedom is to choose $j$ by some procedure other than flipping a fair coin. 
This gives her no advantage, assuming -- as we will here -- 
that Alice and Bob each assign equal utility to Bob knowing the datum at points 
$(x, y, z, t)$ and $(-x, y, z,t)$, for any coordinate values.\footnote{If their utilities are asymmetric,
things become more complicated, and we need game theory to determine Alice's and Bob's optimal actions.}

Moreover, given this symmetry of utilities, it can only help Bob if Alice deviates from
unbiased randomness: if he learns some information about Alice's strategy for
choosing $j$ he may alter his own strategy in response, and thus increase the 
chance that the datum is transferred to him earlier.    
We need not consider this possibility, since it involves both parties violating the
protocol\footnote{As usual in mistrustful cryptography, the aim is to protect an honest party against a cheat; 
we do not aim to protect cheats against the consequences of their own misbehaviour.}, 
but in any case there is no motivation for Alice to pursue this strategy.   

\subsection{Security against Bob}

Bob may act as he wishes, constrained only by the laws of quantum theory and relativity.
By definition, he cannot cheat.   Interestingly, though, he is nonetheless quite constrained. 
We say Bob {\it learns}\footnote{I.e. comes to know with certainty.}
the correct value of $i$ if he obtains it by some strategy
which (in the ideal error-free case) guarantees that, if it produces an answer
$i'$ and Alice chose a classically pre-determined $i$, then $i=i'$.

Suppose first that he follows the strategy above, sending his qudit in a randomly chosen
direction $v_k$ ($k = 0$ or $1$), and suppose without loss of generality that $k=0$.
He then learns the value of $j$ at $Q_0$, where he either receives the second qudit
($j=0$) or does not ($j=1$); he similarly learns this value at $Q_1$.   If $j=0$, he can obtain the datum $i$ at $Q_0$, 
by carrying out a local measurement on both qudits there.    If $j=1$, he now has the two qudits at spacelike
separated locations, $Q_0$ and $Q_1$, and is aware of the situation at both locations. 
For any point $X$, we write $L( X)$ for the set containing $X$ and its future light cone.     
Bob can obtain the datum $i$ at any point $Y$ in the intersection $L(Q_0 ) \cap L(Q_1)$ 
by sending the two qudits from $Q_0$ and $Q_1$ to $Y$.\footnote{
He could also achieve this by a variety of other strategies, for example
by using a second predistributed entangled pair of qudits to carry
out a nonlocal measurement at $Q_0$ and $Q_1$ and sending the
results to $Y$.}   And this is the best Bob can do when he holds the two entangled qudits
at $Q_0$ and $Q_1$: he cannot then 
learn the datum $i$ at any location outside  $L(Q_0 ) \cap L(Q_1)$.    

In summary, following this strategy, with probability $\frac{1}{2}$ he learns $i$ at $Q_j$,
and with probability $\frac{1}{2}$ he learns $i$ only at $Y$.    Note that Alice never learns $k$, and
so does not know the location at which Bob learns $i$, since all Bob's
operations are private.   

In fact, the above strategy is optimal for Bob, in the following sense:
(i) Bob has no strategy
that gives him probability greater than $0$ of learning $i$ at any point outside $L( Q_j )$, and (ii) Bob has
no strategy that gives
him probability greater than $\frac{1}{2}$ of learning $i$ at any point outside $L( Q_0) \cap L(Q_1 )$.  

To see (i), we simply observe that Bob has no access to the subsystem defined by
${\cal C}^d_A$ until Alice transfers it at $Q_j$.  

To see (ii), consider a spacelike hypersurface $S$ close to and in the past of the boundary of 
$L( Q_0) \cup L(Q_1 )$.    For simplicity, take $S$ to be symmetric under reflections in the
$x$-coordinate, and define the subsurfaces $S_0 = \{ (x,y,z,t ) \in S \, : \,  x \leq 0 \}$
and $S_1 = \{ (x,y,z,t)  \in S \, : \,  x \geq 0  \}$. 

Without loss of generality, we can consider Bob's actions between $P$ and $S$ as defined
by a unitary operation acting on  $ {\cal C}^d_B$ together perhaps with ancillary states, generating
some quantum state on $S$.   
Hence at least one of the states defined by the restriction of this state to $S_0$ or
$S_1$ must be entangled with the $ {\cal C}^d_A$ state held by Alice.  
As Bob does not know $j$ until after the points $Q_0$ and $Q_1$, his actions up to 
$S$ are independent of the choice of $j$.
Hence, with probability $\geq \frac{1}{2}$, Bob generates a state on $S_{\bar{j}}$ -- 
the opposite half of the hypersurface to $Q_j$ -- that is entangled with the state transferred by
Alice at $Q_j$.     Bob thus cannot learn the outcome of the desired measurement at any point
not in the intersection of the future light cone of $Q_j$ and the 
future light cone of at least one point in $S_{\bar{j}}$ .     
As we can find spacelike $S$ arbitrarily close to the boundary of $L( Q_0) \cup L(Q_1 )$, 
the result follows. 

\section{Discussion}

The task defined here, transferring data at a location that remains
hidden from the transferrer, has
no precise analogue in classical or non-relativistic quantum cryptography,
although it bears some resemblance to classical oblivious 
transfer \cite{rabinot}.  
Like classical oblivious transfer, it is probably not very useful in
itself, but could be a building block for much more useful
tasks (as oblivious transfer is
\cite{otfounding}).   

In any case, it illustrates some interesting 
possibilities in relativity-based cryptography.   First, one can 
aim to control where exactly
in spacetime various parties learn information as a result of a
protocol.    In our protocol
this control is randomized; it would also be interesting to explore
the (im)possibility of deterministic location control.     Second, 
the locations at which information is learned
themselves constitute information generated by the protocol, and the
flow of this information is also controlled.   Moreover, this control
can be asymmetric -- in our protocol, Bob learns the location information, 
and Alice never does.

These possibilities raise some intriguing questions.   
For example, one can imagine
secure multi-party computation \cite{multi} scenarios in which parties are supposed to learn some 
outputs given by prescribed joint functions of the inputs, but it does not matter that various parties eventually
learn partial or even complete data about other parties' input data, as long as there are strong
enough constraints on where in spacetime they learn this.   It would
be very interesting to understand whether, and under what conditions,
such constraints can be implemented securely.   

Our protocol imposes strong constraints on Bob whatever it does.
However, it should be stressed
that he has more general possibilities than those highlighted above, if he is willing to accept
some risk that he will never obtain the correct value of Alice's input datum.   For example,
he could carry out some form of partial cloning on his qudit and send the two outputs to
$Q_0$ and $Q_1$ respectively, and then obtain some information about $i$ by a local
measurement at $Q_j$, whichever $j$ Alice chooses.    It would be interesting to 
characterize the tradeoffs available to Bob via his most general strategies. 

Finally, one might perhaps wonder whether the protocol is
intrinsically quantum, or whether it could perhaps be implemented
classically.  For example, could Alice initially give Bob a classical
bit $b_0$ at $P$, and a second bit $b_1$ at one of the $Q_i$, in order
to transfer the bit sum $b = b_0 \oplus b_1$?   A little thought shows
this does not restrict Bob, since once he receives $b_0$ he can simply
broadcast it in all directions, ensuring that he obtains $b$ at
whichever point Alice chooses to return $b_1$.  
More generally, since the security proof relies on the no-summoning
theorem, which holds in relativistic quantum theory but not in
relativistic classical theories, no classical protocol can achieve
the same result.

\acknowledgments
This work was partially supported by an FQXi mini-grant and by Perimeter Institute for Theoretical
Physics. Research at Perimeter Institute is supported by the Government of Canada through Industry Canada and
by the Province of Ontario through the Ministry of Research and Innovation.

%\bibliography{../bibtex}

\end{document}